\documentstyle[epsf,12pt]{article}
\newlength{\dinwidth}
\newlength{\dinmargin}
\newcommand{\Pma}{I\!\!P}
\begin{document}
\input epsf
\vspace{1 cm}
\begin{titlepage}

\title {
{\bf Diffractive Hard Photoproduction at HERA and Evidence for the
     Gluon Content of the Pomeron}
     \\
\author{ ZEUS Collaboration } }
\date{ }
\maketitle
\vspace{5 cm}
\begin{abstract}
 Inclusive jet cross sections for events with a large
rapidity gap with respect to the proton direction from the reaction $ep
\rightarrow jet \; + \; X$ with quasi-real photons have been measured
with the ZEUS detector. The cross sections refer to jets with
transverse energies $E_T^{jet}>8$~GeV. The data show the
characteristics of a diffractive process mediated by pomeron exchange.
Assuming that the events are due to the exchange of a pomeron with
partonic structure, the quark and gluon content of the pomeron is
probed at a scale $\sim (E_T^{jet})^2$. A comparison of
the measurements with model predictions based on QCD plus
Regge phenomenology requires a contribution of partons with a hard
momentum density in the pomeron. A combined analysis of the jet cross
sections and recent ZEUS measurements of the diffractive structure
function in deep inelastic scattering gives the first experimental
evidence for the gluon content of the pomeron in diffractive hard
scattering processes. The data indicate that between 30\% and 80\% of
the momentum of the pomeron carried by partons is due to hard gluons.

\end{abstract}
\setcounter{page}{0}
\thispagestyle{empty}   
\eject
\def\3{\ss}
\textwidth 15.5cm
\parindent 0cm
\footnotesize
\renewcommand{\thepage}{\Roman{page}}
\begin{center}
\begin{large}
The ZEUS Collaboration
\end{large}
\end{center}
M.~Derrick, D.~Krakauer, S.~Magill, D.~Mikunas, B.~Musgrave,
J.~Repond, R.~Stanek, R.L.~Talaga, H.~Zhang \\
{\it Argonne National Laboratory, Argonne, IL, USA}~$^{p}$\\[6pt]
R.~Ayad$^1$, G.~Bari, M.~Basile,
L.~Bellagamba, D.~Boscherini, A.~Bruni, G.~Bruni, P.~Bruni, G.~Cara
Romeo, G.~Castellini$^{2}$, M.~Chiarini,
L.~Cifarelli$^{3}$, F.~Cindolo, A.~Contin, M.~Corradi,
I.~Gialas$^{4}$,
P.~Giusti, G.~Iacobucci, G.~Laurenti, G.~Levi, A.~Margotti,
T.~Massam, R.~Nania, C.~Nemoz, \\
F.~Palmonari, A.~Polini, G.~Sartorelli, R.~Timellini, Y.~Zamora
Garcia$^{1}$,
A.~Zichichi \\
{\it University and INFN Bologna, Bologna, Italy}~$^{f}$
% ...
\\[6pt]
\clearpage
\hspace*{1mm}
$^{ 1}$ supported by Worldlab, Lausanne, Switzerland \\
\hspace*{1mm}
$^{ 2}$ also at IROE Florence, Italy  \\
\hspace*{1mm}
$^{ 3}$ now at Univ. of Salerno and INFN Napoli, Italy  \\
\hspace*{1mm}
$^{ 4}$ supported by EU HCM contract ERB-CHRX-CT93-0376 \\
\hspace*{1mm}
% ...
$^{18}$ now with OPAL Collaboration, Faculty of Physics at Univ. of
        Freiburg \\
% ...

\begin{tabular}{lp{15cm}}
$^{a}$ &supported by the Natural Sciences and Engineering Research
         Council of Canada (NSERC) \\
$^{b}$ &supported by the FCAR of Qu\'ebec, Canada\\
$^{c}$ &supported by the German Federal Ministry for Research and
         Technology (BMFT)\\
$^{d}$ &supported by the MINERVA Gesellschaft f\"ur Forschung GmbH,
         and by the Israel Academy of Science \\
$^{e}$ &supported by the German Israeli Foundation, and
         by the Israel Academy of Science \\
$^{f}$ &supported by the Italian National Institute for Nuclear Physics
         (INFN) \\
$^{g}$ &supported by the Japanese Ministry of Education, Science and
         Culture (the Monbusho)
         and its grants for Scientific Research\\
$^{h}$ &supported by the Korean Ministry of Education and Korea Science
         and Engineering Foundation \\
$^{i}$ &supported by the Netherlands Foundation for Research on Matter
         (FOM)\\
$^{j}$ &supported by the Polish State Committee for Scientific Research
         (grant No. SPB/P3/202/93) and the Foundation for Polish-
         German Collaboration (proj. No. 506/92) \\
$^{k}$ &supported by the Polish State Committee for Scientific
         Research (grant No. PB 861/2/91 and No. 2 2372 9102,
         grant No. PB 2 2376 9102 and No. PB 2 0092 9101) \\
$^{l}$ &partially supported by the German Federal Ministry for
         Research and Technology (BMFT) \\
$^{m}$ &supported by the German Federal Ministry for Research and
         Technology (BMFT), the Volkswagen Foundation, and the Deutsche
         Forschungsgemeinschaft \\
$^{n}$ &supported by the Spanish Ministry of Education and Science
         through funds provided by CICYT \\
$^{o}$ &supported by the Particle Physics and Astronomy Research
        Council \\
$^{p}$ &supported by the US Department of Energy \\
$^{q}$ &supported by the US National Science Foundation
\end{tabular}
\end{titlepage}
\newpage
\section{\bf Introduction}

 Electron-proton collisions at HERA have shown evidence for hard
processes in diffractive reactions. Both in deep inelastic scattering
(DIS) ($Q^2 > 10$~GeV$^2$, where $Q^2$ is the virtuality of the
exchanged photon) \cite{zelrgd93,zelrgd94,h1lrgd94} and in
photoproduction ($Q^2 \sim 0$) \cite{zelrgp94,h1lrgp94}, events
characterized by a large rapidity gap towards the proton direction
have been observed and interpreted as resulting from diffractive
scattering \cite{zelrgd93,zelrgd94,zelrgp94}. In the DIS regime, hard
scattering for this class of events has been revealed through the
virtuality of the probing photon \cite{zelrgd93,h1lrgd94} and through
the observation of jet structure in the final state \cite{zelrgd94}. In
the photoproduction domain, the hard scattering has been identified
through jet production \cite{zelrgp94,h1lrgp94}.

 Diffractive processes are generally considered to proceed through the
exchange of a colourless object with the quantum numbers of the
vacuum, generically called the pomeron ($\Pma$). Although the
description of soft diffractive processes in terms of pomeron exchange
has been a phenomenological success, the description of the pomeron in
terms of a parton structure at first lacked experimental support. On
the basis of $pp$ data \cite{r608} from the CERN ISR, Ingelman and
Schlein \cite{ingelsch} suggested that the pomeron may have a partonic
structure. The observation of jet production in $p\bar{p}$ collisions
with a tagged proton (or antiproton) made by the UA8 Collaboration
\cite{ua8coll} gave strong evidence for such a structure. Further
evidence has been provided by the observations made at HERA $[1-5]$,
which in addition include the first measurements of the diffractive
structure function in DIS \cite{h1lrgd95,zelrgd95}.

 The description of diffractive processes in terms of QCD has
remained elusive, in part due to the lack of a sufficiently large
momentum transfer on which to base the perturbative expansion.
Cross sections for diffractive processes involving large transverse
energy jets or leptons in the final state are, however, amenable to
perturbative QCD calculations $[7,11-17]$. Their measurement could
answer several questions concerning the structure of the pomeron such
as: whether the parton picture is valid for the pomeron and universal
parton densities can be defined; what fraction of the pomeron momentum
is carried by gluons and what by quarks; and whether a momentum sum
rule applies to the pomeron \cite{capella}.

 This paper presents the first measurement of inclusive jet cross
sections in photoproduction at centre-of-mass energies $\sim 200$~GeV
with a large rapidity gap.
This process is sensitive to both the gluon and quark content
of the pomeron. In order to examine the partonic structure of the
pomeron, these jet cross sections are compared to predictions from
models based on perturbative QCD and Regge phenomenology. The jet cross
sections measured in photoproduction, combined with the results on the
diffractive structure function in deep inelastic scattering
\cite{zelrgd95}, give the first experimental evidence for the gluon
content of the pomeron. The result does not depend on the flux of
pomerons from the proton nor on the assumption that a momentum sum
rule can be defined for the pomeron. The data sample used in this
analysis corresponds to an integrated luminosity of 0.55~pb$^{-1}$ and
was collected during 1993 with the ZEUS detector at HERA.

\section{Experimental setup}

\subsection{HERA operation}

The experiment was performed at the electron-proton collider HERA
using the ZEUS detector. During 1993 HERA operated with electrons of
energy $E_e=26.7$ GeV colliding with protons of energy $E_p=820$ GeV.
HERA is designed to run with 210 bunches separated by 96 ns in each of
the electron and proton rings. For the 1993 data-taking, 84 bunches
were filled for each beam  and an additional 10 electron and 6
proton bunches were left unpaired for background studies. The electron
and proton beam currents were typically 10 mA, with instantaneous
luminosities of approximately $ 6 \cdot 10^{29}$~cm$^{-2}$~s$^{-1}$.

\subsection{The ZEUS detector and trigger conditions}

 ZEUS is a multipurpose magnetic detector. The configuration for the
1993 running period has been described elsewhere
\cite{zelrgd94,status}. A brief description concentrating on those
parts of the detector relevant to this analysis is presented here.

Charged particles are tracked by two concentric cylindrical drift
chambers, the vertex detector (VXD) and the central tracking detector
(CTD), operating in a magnetic field of 1.43~T provided by a thin
superconducting coil. The coil is surrounded by a high-resolution
uranium-scintillator calorimeter (CAL) divided into three parts,
forward\footnote{The ZEUS coordinate system is defined as right-handed
with the $Z$ axis pointing in the proton beam direction, hereafter
referred to as forward, and the $X$ axis horizontal, pointing towards
the centre of HERA.} (FCAL) covering the pseudora- pidity\footnote{The
pseudorapidity is defined as $-\ln(\tan\frac{\theta}{2})$, where the
polar angle $\theta$ is taken with respect to the proton beam
direction, and is denoted by $\eta_d$ ($\eta$) when the polar angle is
measured with respect to the nominal interaction point (the
reconstructed vertex of the interaction).} region\footnote{The FCAL has
the forward edge at $\eta_{\rm edge}=4.3$ with full acceptance for
$\eta_d <3.7$.} $4.3\geq \eta_d \geq 1.1$, barrel (BCAL) covering the
central region $1.1 \geq \eta_d \geq -0.75$ and rear (RCAL) covering
the backward region $-0.75\geq \eta_d \geq -3.8$. The solid angle
coverage is $99.7\%$ of $4\pi$.  The CAL parts are subdivided into
towers which in turn are subdivided longitudinally into electromagnetic
(EMC) and hadronic (HAC) sections. The sections are subdivided into
cells, each viewed by two photomultiplier tubes. The CAL is
compensating,  with equal response
to hadrons and electrons. Measurements under test beam conditions show
that the energy resolution is $\sigma_E/E = 0.18/\sqrt{E}$ ($E$ in GeV)
for electrons and $\sigma_E/E = 0.35/\sqrt{E}$ for hadrons
\cite{calori}. In the analysis presented here, CAL cells with EMC (HAC)
energy below 60~MeV (110~MeV) are excluded to minimize the effect of
calorimeter noise. This noise is dominated by the uranium activity and
has an r.m.s. value below 19~MeV for EMC cells and below 30~MeV for HAC
cells. For measuring the luminosity as well as for tagging very small
$Q^2$ processes, two lead-scintillator calorimeters \cite{lumi},
located at 107~m and 35~m downstream from the interaction point in the
electron direction, detect the bremsstrahlung photons and the
scattered electrons respectively.

Data were collected using a three-level trigger \cite{status}. The
first-level trigger (FLT) is built as a deadtime-free pipeline. The
FLT for the sample of events analysed in this paper required a logical
OR of different conditions on sums of energy in the CAL cells. The
average FLT acceptance for the events under study was approximately
90\%. The second-level trigger used information from a subset of
detector components to differentiate physics events from backgrounds
consisting mostly of proton beam gas interactions. The third-level
trigger (TLT) used the full event information to apply specific
physics selections. For this analysis, the following conditions
were required: a) the event has a vertex reconstructed by the tracking
chambers (VXD+CTD) with the $Z$ value in the range $|Z| <$~75~cm;
b) $E - p_Z \geq$ 8~GeV, where $E$ is the total energy as measured by
the CAL
    $$ E = \sum_i E_i ,$$
$p_Z$ is the $Z$-component of the vector
    $$\vec p=\sum_i E_i \vec r_i ,$$
the sums run over all CAL cells, $E_i$ is the energy of the
calorimeter cell $i$  and $\vec r_i$ is a unit vector along the line
joining the reconstructed vertex and the geometric centre of the cell
$i$; c) $p_Z/E\leq 0.94$ to reject beam-gas interactions; and
d) the total transverse energy as measured by the CAL, excluding the
cells whose polar angles are below $10^{\circ}$, exceeds 12~GeV.

\section{\bf Diffractive hard photoproduction}

 Diffractive hard photoproduction processes in $ep$ collisions are
characterized by $Q^2 \approx 0$ and by a final state consisting of a
hadronic system $X$ containing one or more jets, the scattered electron
and the scattered proton
\begin{equation}
 e + p_i \rightarrow e + X + p_f \rightarrow e + ( jet + X_r ) + p_f
\end{equation}
where $p_i$ ($p_f$) denotes the initial (final) state proton and
$X$ consists of at least one jet plus the remaining hadronic system
($X_r$).

 The kinematics of this process are described in terms of four
variables. Two of them describe the electron-photon vertex and can be
taken to be the virtuality of the exchanged photon ($Q^2$) and the
inelasticity variable $y$ defined by
$$ y=1-\frac{E_e^{\prime}}{E_e} \frac{1-\cos{\theta_e^{\prime}}}{2}$$
where $E_e^{\prime}$ denotes the scattered electron energy and
$\theta_e^{\prime}$ is the electron scattering angle. The other
two variables describe the proton vertex: the fraction of the momentum
of the initial proton carried by the scattered proton ($x_f$), and the
square of the momentum transfer ($t$) between the initial and final
state proton.  In terms of these variables and at low values of $Q^2$
and $t$, the square of the mass of the hadronic system $X$ is given by
\begin{equation}
   M^2_X \approx (1-x_f) \; y \; s
\end{equation}
where $s$ is the square of the $ep$ centre-of-mass energy.

 Diffractive processes in which the photon dissociates give rise to a
large rapidity gap between the
hadronic system $X$ and the scattered proton:
\begin{equation}
  \Delta y_{GAP} = y_{p_f} - y^{had}_{max}
\end{equation}
where $y_{p_f}$ is the rapidity of the scattered proton and
$y^{had}_{max}$ is the rapidity of the most forward going hadron
belonging to the
system $X$. The same signature is expected for double dissociation
where the scattered proton is replaced by a low mass baryonic system
($N$). In this paper, the outgoing proton (or system $N$) was not
observed, and instead of $y^{had}_{max}$ the pseudorapidity
($\eta^{had}_{max}$) of the most forward-going hadron in the detector
was used.

 Two cross sections are presented in this paper. First, the cross
section for inclusive jet production is measured as a function of the
pseudorapidity of the jet ($\eta^{jet}$) (for the definition of the
jet variables see section~4) in reaction~(1) with the most-forward
going hadron at $\eta^{had}_{max} < 1.8$. This corresponds to a
rapidity gap of at least 2.5 units measured from the edge of the CAL
($\Delta\eta_{GAP} = \eta_{\rm edge} - \eta^{had}_{max}$). This cross
section is denoted by
\begin{equation}
     \frac{d\sigma}{d\eta^{jet}}(\eta^{had}_{max}<1.8)
\end{equation}
and is measured in the $\eta^{jet}$ range between $-1$ and
$1$. Second, the integrated cross section for inclusive jet production
is determined as a function of $\eta^{0}_{max}$
\begin{equation}
   \sigma(\eta^{had}_{max}<
          \eta^{0}_{max}) = \int_{-1}^{+1} d\eta^{jet}
   \frac{d\sigma}{d\eta^{jet}}(\eta^{had}_{max}<\eta^{0}_{max})
\end{equation}
and is measured in the range of $\eta^{0}_{max}$ between $1$ and $2.4$.
Both measurements include contributions from double dissociation where
the large-rapidity-gap requirement is satisfied.

 The jet cross sections refer to jets at the hadron level with
a cone radius, $R = \sqrt{\Delta\eta^2+\Delta\phi^2}$, of one unit in
pseudorapidity ($\eta$) $-$ azimuth ($\phi$) space and integrated over
the transverse energy of the jet $E^{jet}_T>8$~GeV. They are given in
the kinematic region $Q^2 < 4$~GeV$^2$ and $0.2 < y < 0.85$. This
region corresponds to photoproduction interactions at centre-of-mass
energies in the range 130-270~GeV with a median $Q^2 \approx
10^{-3}$~GeV$^2$.

\subsection{\bf Models}

 The description of diffractive hard processes in terms of QCD is
still in an early stage. Two main theoretical approaches have been
considered. Both assume that a pomeron ($\Pma$) is emitted by the
proton. The variable $x_{\Pma}\equiv 1-x_f$ is then the fraction of the
initial proton's momentum carried by the pomeron and $M^2_X \approx
x_{\Pma} y s$ is the square of the $\gamma \Pma$ centre-of-mass energy.
The two approaches differ in the modelling of jet production in
$\gamma \Pma$ collisions. One of them $[7,11-13]$ assumes
factorisation (factorisable models) while the other one does not
$[14,16]$ (non-factorisable models). The latter are, however, not
considered in what follows due to the lack of a Monte Carlo generator
with an appropriate description of the event jet structure.

 Calculations based on factorisable models involve three basic
ingredients: the flux of pomerons from the proton as a function of
$x_{\Pma}$ and $t$, the parton densities in the pomeron and the matrix
elements for jet production. The pomeron is assumed to be a source of
partons which interact either with the photon (direct component) or
with a partonic constituent of the photon (resolved component).
As an example, the contribution of the direct component to the cross
section for reaction~(1) is given by
\begin{equation}
 \sigma_{dir} = \int dy f_{\gamma/e}(y) \int\int dx_{\Pma} dt
  f_{\Pma/p}(x_{\Pma},t) \sum_i \int d\beta
  \sum_{j,k} \int d\hat{p}^2_T
  \frac{d\hat{\sigma}_{i+\gamma \rightarrow j + k }}{
  d\hat{p}^2_T}(\hat{s},\hat{p}^2_T,\mu^2) \; \;
  f_{i/\Pma}(\beta,\mu^2)
\end{equation}
where $f_{\gamma/e}$ is the flux of photons from the
electron\footnote{ The $Q^2$ dependence has been integrated out using
the Weizs\"{a}cker-Williams approximation.} and $f_{\Pma/p}$ is the
flux of pomerons from the proton. The sum in $i$
runs over all possible types of partons present in the pomeron, and
$f_{i/\Pma}(\beta,\mu^2)$ is the density of partons of type $i$
carrying a fraction $\beta$ of the pomeron momentum at a scale
$\mu^2$ and is assumed to be independent of $t$. The sum in $j$ and $k$
runs over all possible types of final state partons and
$\hat{\sigma}_{i+\gamma \rightarrow j+k}$ is the cross section for the
two-body collision $i + \gamma \rightarrow j + k$ and depends on
the square of the centre-of-mass energy ($\hat{s}$), the transverse
momentum of the two outgoing partons ($\hat{p}_T$) and the momentum
scale ($\mu$) at which the strong coupling constant ($\alpha_s(\mu^2)$)
is evaluated. One possible choice is $\mu^2 = \hat{p}^2_T$. In these
models, the pomeron flux factor is extracted from hadron-hadron
collisions using Regge theory, and the matrix elements are computed in
perturbative QCD. However, the parton densities in the pomeron have to
be extracted from experiment. While the recent measurements of the
diffractive structure function in DIS at HERA \cite{h1lrgd95,zelrgd95}
give information on the quark densities in the pomeron, the gluon
content has so far not been determined.

 Two forms of the pomeron flux factor are commonly used. The
Ingelman-Schlein form (IS) \cite{pompyt} uses a parametrisation of
UA4 data \cite{ua4}:
\begin{equation}
 f_{\Pma/p}(x_{\Pma},t)=\frac{c_0}{x_{\Pma}} \cdot
  ( 3.19 e^{8 t} + 0.212 e^{3 t} )
\end{equation}
where $c_0 = \frac{1}{2.3}$~GeV$^{-2}$ and $t$ is in GeV$^2$.
The Donnachie-Landshoff form (DL) \cite{donlan} is calculated in Regge
theory, with parameters determined by fits to hadron-hadron data:
\begin{equation}
 f_{\Pma/p}(x_{\Pma},t) = \frac{9 b_0^2}{4 \pi^2} F_1(t)^2
   x_{\Pma}^{1-2\alpha(t)}
\end{equation}
using the elastic form factor $F_1(t)$ of the proton, the pomeron-quark
coupling $b_0 \simeq 1.8$~GeV$^{-1}$ and the pomeron trajectory
$\alpha(t) = 1.085 + 0.25 t$ with $t$ in GeV$^2$.

 Various parametrisations of the parton densities in the pomeron
have been suggested on theoretical grounds $[7,11-13]$. The following
represent extreme possibilities for the shape of the quark and gluon
momentum densities:
\begin{itemize}
\item hard gluon density
      $\beta f_{g/\Pma}(\beta,\mu^2)=6\beta(1-\beta)$;
\item soft gluon density $\beta f_{g/\Pma}(\beta,\mu^2)=6(1-\beta)^5$;
\item hard quark density (for two flavours)
    $\beta f_{q/\Pma}(\beta,\mu^2)= \frac{6}{4} \beta(1-\beta)$.
\end{itemize}
The first two assume a pomeron made entirely of gluons and the last one
a pomeron made of $u\bar{u}$ and $d\bar{d}$ pairs. In all cases a
possible $\mu^2$ dependence of the parton densities is
neglected\footnote{The $\mu^2$ dependence of the parton densities
in the pomeron is expected to be smaller than the differences between
the various parton densities considered \cite{ingelsch}.} and the
densities are normalised such that all of the pomeron's momentum is
carried by the partons under consideration, $\Sigma_{\Pma}(\mu^2)
\equiv \int_0^1 d\beta \sum_i \beta f_{i/\Pma}(\beta,\mu^2)=1$.

 However, since the pomeron is not a particle, it is unclear whether
or not the normalisations of the pomeron flux factor and the momentum
sum of the pomeron can be defined independently. Nevertheless, for an
assumed normalisation of the flux factor, the momentum sum
$\Sigma_{\Pma}(\mu^2)$ can be measured. The definition used for the
DL form of the pomeron flux factor is the appropriate one if the
pomeron were an ordinary hadron and, hence, the one in which the
momentum sum rule might be fulfilled \cite{donlan87}.

 Factorisable models presently account only for diffractive hard
processes in which the proton remains intact. Since the measurements
are based on the requirement of a large rapidity gap in the central
detector, the contribution to the measured cross sections from
double dissociation has to be taken into account when
comparing with model predictions.

\section{Data selection and jet search}

 Events from quasi-real photon proton collisions were selected
using the same criteria as reported earlier \cite{zesep94}. The main
steps are briefly discussed here.

 Events satisfying the TLT selection described in section 2.2 are
first selected. A cone algorithm in $\eta$-$\phi$ space with a cone
radius of $1$ unit \cite{cone,snow} is then used to reconstruct jets,
for both data and simulated events (see next section) from the energy
deposits in the CAL cells ($cal$ jets), and for simulated events also
from the final state
hadrons ($had$ jets). The axis of the jet is defined according to the
Snowmass convention \cite{snow}, $\eta^{jet}$ ($\phi^{jet}$) is the
transverse energy weighted mean pseudorapidity (azimuth) of all the
objects (CAL cells or final state hadrons) belonging to that jet. This
procedure is explained in detail elsewhere \cite{zesep94}. The
variables associated with the $cal$ jets are denoted by
$E^{jet}_{T,cal}$, $\eta_{cal}^{jet}$, and $\phi_{cal}^{jet}$, while
the ones for the $had$ jets by $E^{jet}_{T}$, $\eta^{jet}$, and
$\phi^{jet}$.

 A search for jet structure using the CAL cells is performed in the
data. Events with at least one jet fulfilling the conditions
$E^{jet}_{T,cal} > 6$~GeV and $-1 < \eta^{jet}_{cal} < 2$ are retained.
Beam-gas interactions, cosmic-ray showers, halo muons and DIS neutral
current events are removed from the sample as described previously
\cite{zesep94}. The sample thus consists of events from $ep$
interactions with $Q^2 < 4$~GeV$^2$ and a median $Q^2 \approx
10^{-3}$~GeV$^2$. The $\gamma p$ centre-of-mass energy ($W$) is
calculated using the expression $W=\sqrt{y s}$. The event
sample is restricted to the kinematic range $0.2 < y < 0.85$ using the
following procedure. The method of Jacquet-Blondel \cite{jacblo} is
used to estimate $y$ from the energies measured in the CAL cells
(see section 2.2)
       $$ y_{JB} = \frac{E - p_Z}{2 E_e}.$$
As can be verified using photoproduction events with an electron
detected in the luminosity monitor (tagged events), $y_{JB}$
systematically underestimates $y$ by approximately 20\%, which
is adequately reproduced in the Monte Carlo simulation of the
detector. To allow for this effect, the event selection required
$0.16 < y_{JB} < 0.7$. The sample thus obtained consists of 19,485
events containing 24,504 jets. The only significant background, which
is from misidentified DIS neutral current interactions with $Q^2 >
4$~GeV$^2$, is estimated to be below 2\%. The photoproduction origin of
the sample is verified by the expected contribution (26\%) of tagged
events.

\section{\bf Monte Carlo simulation}

 Events from diffractive hard photoproduction processes were simulated
using the program POMPYT \cite{pompyt}. These events were used to
determine both the response of the detector to the hadronic final state
and the correction factors for the cross sections for jet production
with a large rapidity gap.

 The POMPYT generator is a Monte Carlo implementation of the model
proposed in \cite{ingelsch}. The generator makes use of the program
PYTHIA \cite{pythia} to simulate electron-pomeron interactions via
resolved and direct photon processes. In PYTHIA, the lepton-photon
vertex is modelled according to the Weizs\"{a}cker-Williams
approximation and the effects of initial state bremsstrahlung from the
electron are simulated by using the next-to-leading order electron
structure function \cite{kleiss}. Radiative corrections in our
kinematic region, where $W$ is larger than 100~GeV, are expected to be
negligible \cite{sigtot}. For the resolved processes, the parton
densities of the photon were parametrised according to GS-HO \cite{gs}
and evaluated at the momentum scale set by the transverse momentum of
the two outgoing partons, $\mu^2 = \hat{p}_T^2$. The parton densities
in the pomeron were parametrised according to the forms described in
section~3.1 and were taken to be independent of any scale. In PYTHIA,
the partonic processes are simulated using leading order matrix
elements, with the inclusion of initial and final state parton showers.
Fragmentation into hadrons was performed using the Lund string model
\cite{lund} as implemented in JETSET \cite{jetset}. Samples of events
were generated with different values of the minimum cutoff for the
transverse momentum of the two outgoing partons, starting at
$\hat{p}_{Tmin}= 3$~GeV.

 The program PYTHIA was also used to simulate standard
(non-diffractive) hard photoproduction events via resolved and direct
photon processes. Events were generated using the leading order
predictions of GRV \cite{grv} for the photon parton densities     and
MRSD$_-$ \cite{mrsd} for the proton parton densities.

 All generated events were passed through the ZEUS detector and trigger
simulation programs. They were reconstructed using the same standard
ZEUS off-line programs as for the data.

\section{\bf Event characteristics}

 The event variable $\eta_{max}$, as in previous studies by ZEUS
\cite{zelrgd93,zelrgd94,zelrgp94,zelrgd95}, was used to select events
with a large rapidity gap. For the data, this variable is defined as
the pseudorapidity ($\eta^{cal}_{max}$) of the most forward condensate
with an energy above 400~MeV. A condensate is a contiguous energy
deposit above 100~MeV for pure EMC and 200~MeV for HAC or mixed energy
deposits in CAL. In the samples of simulated events, the $\eta_{max}$
variable is defined at both the hadron and CAL levels. At the hadron
level, all particles with lifetimes larger than $10^{-13}$~s, energies
in excess of 400~MeV and pseudorapidities below 4.5 are considered as
candidates for the most forward final state particle, and
$\eta^{had}_{max}$ defines the pseudorapidity of the most forward
particle. The CAL level uses the same definition as for the data.

\begin{figure}
\epsfysize=18cm
\epsffile{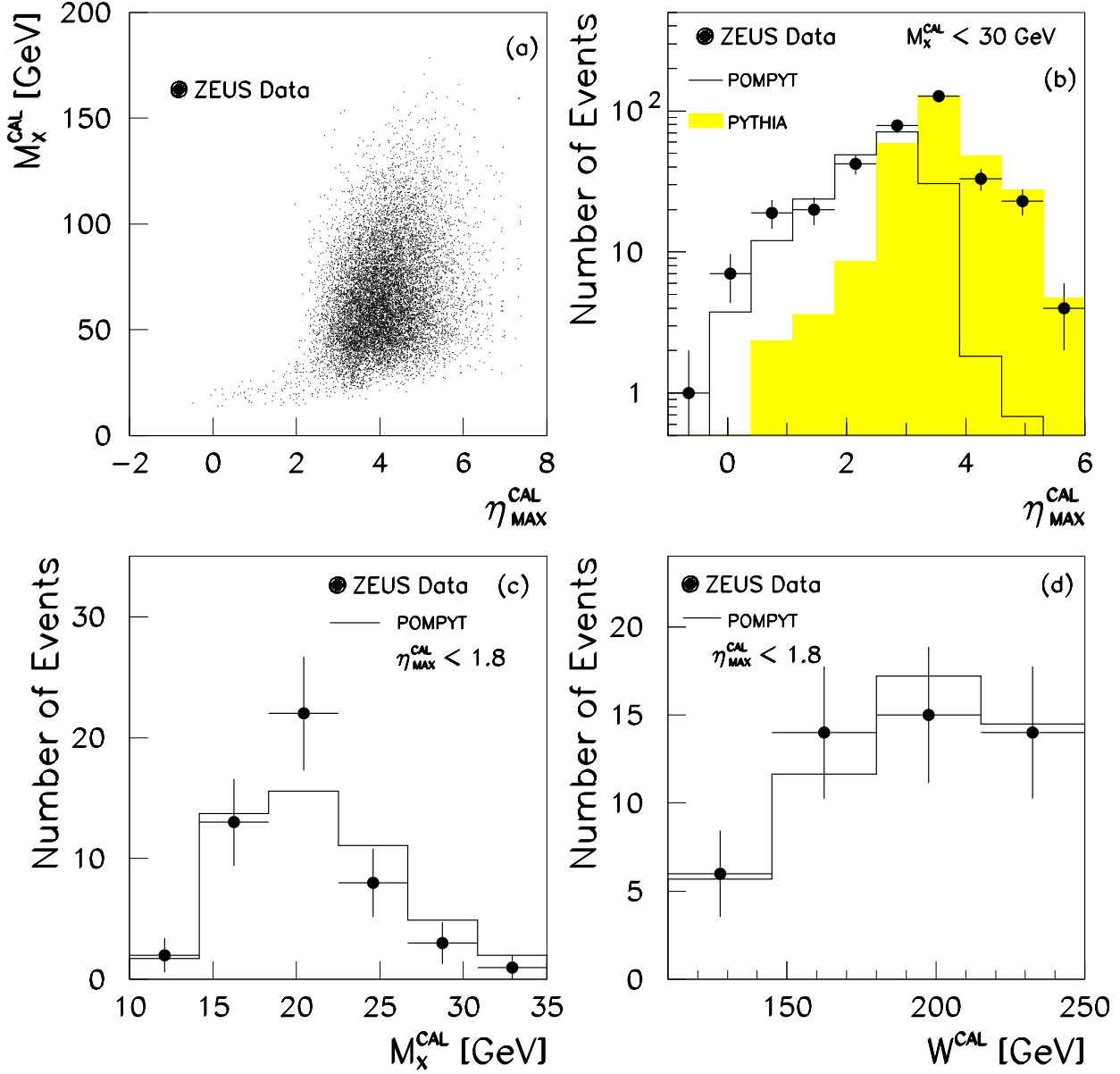}
\caption{\label{figdra0b}{ (a) The scatter plot of $M^{cal}_X$
  versus $\eta^{cal}_{max}$ for the sample of events with at least one
  $cal$ jet fulfilling the conditions $E^{jet}_{T,cal} > 6$~GeV and
  $-1<\eta^{jet}_{cal}<1$; (b) the distribution of $\eta^{cal}_{max}$
  for the events with $M^{cal}_X < 30$~GeV along with the predictions
  of PYTHIA (shaded area) and POMPYT with a hard gluon density in the
  pomeron (solid line). The predictions are normalised to the number of
  data events above and below $\eta^{cal}_{max} = 2.5$, respectively;
  (c) the distribution in $M^{cal}_X$ for the events with
  $\eta^{cal}_{max}<1.8$ together with the prediction of POMPYT with a
  hard gluon density in the pomeron (solid line) normalised to the
  number of data events; (d) the distribution in $W^{cal}$ for the
  events with $\eta^{cal}_{max} < 1.8$ and the
  prediction of POMPYT as in (c).}}
\end{figure}

 The mass of the hadronic system ($M_X$) of each event is reconstructed
using the CAL cells, $M^{cal}_X =\sqrt{E^2 - \vec{p}^2}$
\cite{zelrgp94}. The correlation between $M^{cal}_X$ and
$\eta^{cal}_{max}$ for the sample of events with at least one $cal$
jet fulfilling the conditions $E^{jet}_{T,cal}>6$~GeV and
$-1<\eta^{jet}_{cal}<1$ is displayed in Fig.~\ref{figdra0b}a.
As shown in our previous publication \cite{zelrgp94}, there exists a
distinct class of events with low $\eta^{cal}_{max}$ values. The
large-rapidity-gap events ($\eta^{cal}_{max}<1.8$) are found to
populate the region of low $M^{cal}_X$ values, in contrast to the bulk
of the data which have large $M^{cal}_X$ values. These features of the
data are reproduced by the Monte Carlo simulations: the events from a
simulation of standard hard photoproduction processes using PYTHIA
populate the region of large $\eta^{cal}_{max}$ and large $M^{cal}_X$
values; the events from a simulation of diffractive hard processes
using POMPYT extend into the region of low $\eta^{cal}_{max}$ and
$M^{cal}_X$ values.

 A study of the region of low $M^{cal}_X$ in the data sample reveals
the following features. The $\eta^{cal}_{max}$ distribution for events
with $M^{cal}_X<30$~GeV is shown in Fig.~\ref{figdra0b}b along with the
predictions of PYTHIA and of POMPYT with a pomeron made of hard gluons
(normalised to the number of data events above and below
$\eta^{cal}_{max} = 2.5$, respectively). The simulation of
non-diffractive processes by PYTHIA cannot reproduce the shape of the
measured $\eta^{cal}_{max}$ distribution. On the other hand, the
predictions of POMPYT describe well the shape of the data below
$\eta^{cal}_{max} \sim 3$.

 The $M^{cal}_X$ distribution for the sample of events with
$\eta^{cal}_{max}<1.8$ is shown in Fig.~\ref{figdra0b}c. This sample
consists of 49 events containing 68 jets. The data exhibit an
enhancement at low masses, $15$~GeV$\stackrel{<}{\sim}
M^{cal}_X \stackrel{<}{\sim} 30$~GeV, which is reproduced by
the simulation of POMPYT with a hard gluon density (normalised to
the number of data events). The $W$ of each event is reconstructed
using $y_{JB}$, $W^{cal}= \sqrt{y_{JB} s}$. The distribution of
$W^{cal}$ for events with $\eta^{cal}_{max} <1.8$ is shown in
Fig.~\ref{figdra0b}d along with the expectations of POMPYT with a hard
gluon density (normalised to the number of data events). The $W^{cal}$
dependence exhibited by the data sample is well reproduced by the
simulation of POMPYT. The expectations of POMPYT using a hard quark
density (not shown) also give a good description of the distributions
of the data. Note that POMPYT assumes the cross section
to be independent of $W$ as expected for diffractive processes mediated
by pomeron exchange. The good agreement with the data gives evidence
for the diffractive nature of the large-rapidity-gap events.

 In summary, the data exhibit a different behaviour in the region of
low masses of the hadronic system compared to that of high masses. At
low masses, the shape of the $\eta^{cal}_{max}$ distribution in the
data sample cannot be accounted for by the simulation of
non-diffractive processes as in PYTHIA. On the other hand, the
features of the data are described by the predictions of diffractive
processes mediated by pomeron exchange as in POMPYT. These facts
support the interpretation of these large-rapidity-gap events as being
produced by diffractive processes via pomeron exchange. Therefore, the
measurements of jet cross sections presented in the next section are
compared to the predictions of models based on pomeron exchange.
However, without a detected fast proton in the forward direction, the
jet cross sections refer to events with a large rapidity gap. These
include events with a diffractively
scattered proton as well as those with a diffractively dissociated
proton with mass less than approximately $4$~GeV \cite{zelrgd95}. In
this way, the measurements are presented in a model independent form
suitable for comparison with calculations other than those presented
here.

\subsection{\bf Energy and acceptance corrections}

 The method to correct the transverse energy of a jet as reconstructed
using the CAL cells has been discussed elsewhere \cite{zesep94}. For
samples of simulated events, the transverse energy of a jet as measured
by the CAL ($E^{jet}_{T,cal}$) was compared to that reconstructed using
the final state hadrons ($E^{jet}_{T}$). The corrections to the jet
transverse energy were constructed as multiplicative factors,
$C(E^{jet}_{T,cal},\eta^{jet}_{cal})$, which, when applied to the $E_T$
of the $cal$ jets, give the corrected transverse energies of the jets:
$E^{jet}_{T} = C(E^{jet}_{T,cal},\eta^{jet}_{cal}) \times
E^{jet}_{T,cal}$. The function $C$ corrects for energy losses, and
for values $E^{jet}_{T,cal} > 10$~GeV is approximately flat as a
function of $E^{jet}_{T,cal}$ and varies between $1.08$ and $1.18$
depending on $\eta^{jet}_{cal}$. For $E^{jet}_{T,cal}$ near threshold,
$E^{jet}_{T,cal} \approx $~6~GeV, this correction procedure can give
values as large as 1.40. No correction is needed for
$\eta^{jet}$ ($\eta^{jet} \approx \eta_{cal}^{jet}$). The procedure was
validated by comparing the momenta of the tracks in the $cal$ jet in
data and in Monte Carlo simulations. From this comparison it was
concluded that the energy scale of the jets is corrected to within
$\pm$~5\% \cite{zesep94}. The correction procedure was applied to the
data sample of jets with $E^{jet}_{T,cal}> 6$~GeV to select for further
study those jets with corrected transverse energies of $E^{jet}_T >
8$~~GeV and with the jet pseudorapidity in the range between $-1$ and
1.

 The events generated by POMPYT were used to compute the
acceptance correction for the inclusive jet distributions. This
correction function takes into account the efficiency of the trigger,
the selection criteria and the purity and efficiency of the jet and
$\eta^{had}_{max}$ selection. It also corrects for the migrations in
the variable $\eta^{cal}_{max}$ and yields cross sections for the true
rapidity gap defined by $\eta^{had}_{max}$ and $\eta = 4.5$. After
applying the jet transverse energy corrections, the purity was $\sim
40$\% and the efficiency was $\sim 50$\%. Cross sections were then
obtained by applying bin-by-bin corrections to the inclusive jet
distributions of the data sample in the variables $\eta^{jet}$
and $\eta^{had}_{max}$. The acceptance correction factors for the
inclusive jet cross section $d\sigma/d\eta^{jet}(\eta^{had}_{max}<1.8)$
($\sigma(\eta^{had}_{max}<\eta^0_{max})$) were found to vary between
0.63 and 0.93 (0.59 and 0.84). The dependence of these correction
factors on the choice of parametrisations of the parton densities in
the pomeron were found to be below $\sim 20$\%, and are taken into
account in the systematic uncertainty assigned to the measurements
reported in the next section.

\section{\bf Results}

 In this section, first the measured jet cross sections are presented
and the uncertanties of the measurements discussed. These results
are model-independent. Second, the expectations from
non-diffractive processes are found not to account for the
measurements. Third, the predictions from diffractive models are
compared to the data and estimates of the momentum sum of the pomeron
and of the relative contribution of  quarks and gluons in the pomeron
are extracted using solely the diffractive jet measurements. Fourth,
the jet cross sections in photoproduction are combined with the
measurements of the DIS diffractive structure function to
constrain further the parton content of the pomeron.

\subsection{\bf Jet Cross Sections}

\begin{figure}
\epsfysize=18cm
\epsffile{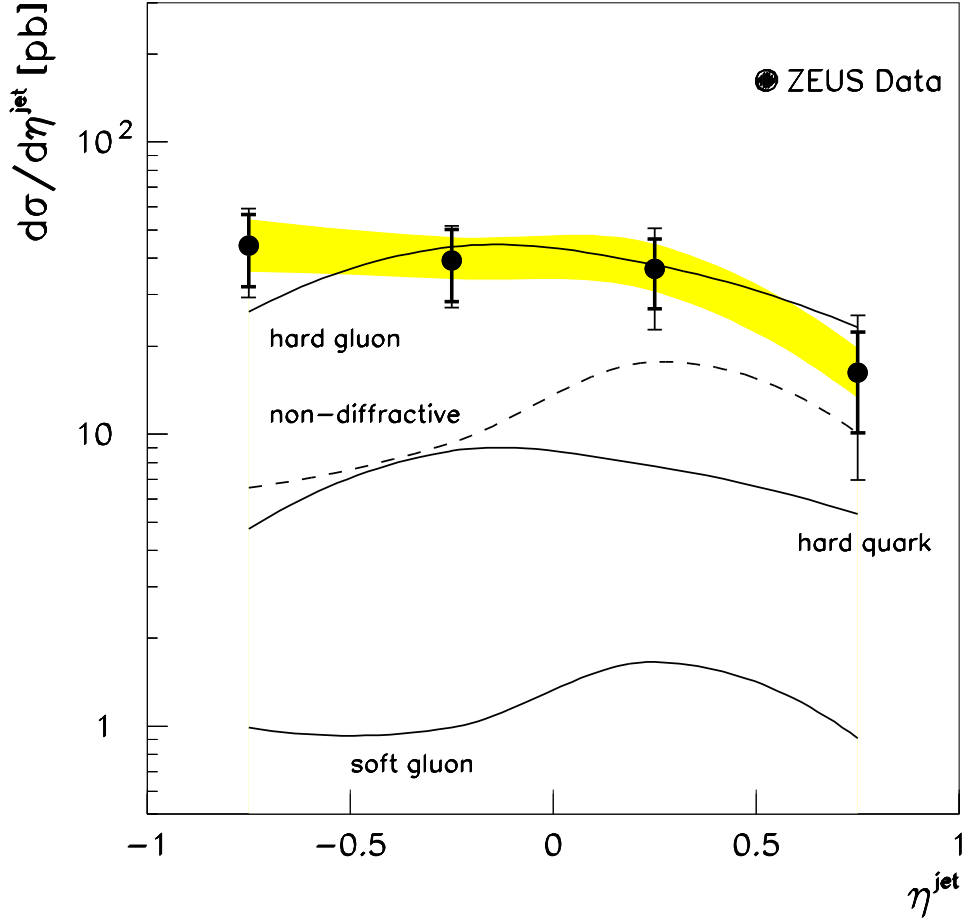}
\caption{\label{figdra1}{ Measured differential $ep$ cross section
  $d\sigma/d\eta^{jet}(\eta^{had}_{max}<1.8)$ for inclusive jet
  production for $E_T^{jet} >$~8~GeV in the kinematic region
  $Q^2 \leq$~4~GeV$^2$ and $ 0.2 < y < 0.85$ (dots). The measurements
  are not corrected for the contributions from non-diffractive
  processes and double dissociation. The inner error bars
  represent the statistical errors of the data, and the total error
  bars show the statistical and systematic errors $-$not associated
  with the energy scale of the jets$-$ added in quadrature. The shaded
  band displays the uncertainty due to the energy scale of the jets.
  For comparison, POMPYT predictions for single diffractive jet
  production ($e+p \rightarrow e+p+jet+X_r$) using the DL flux factor
  for direct plus resolved processes for various
  parametrisations of the pomeron parton densities (hard gluon, upper
  solid line; hard
  quark, middle solid line; soft gluon, lower solid line) are also
  shown. The GS-HO photon parton densities have been used in POMPYT.
  The contribution from non-diffractive processes is exemplified by the
  PYTHIA predictions using MRSD$_-$ (GRV-HO) for the proton (photon)
  parton densities (dashed line).}}
\end{figure}

\begin{figure}
\epsfysize=18cm
\epsffile{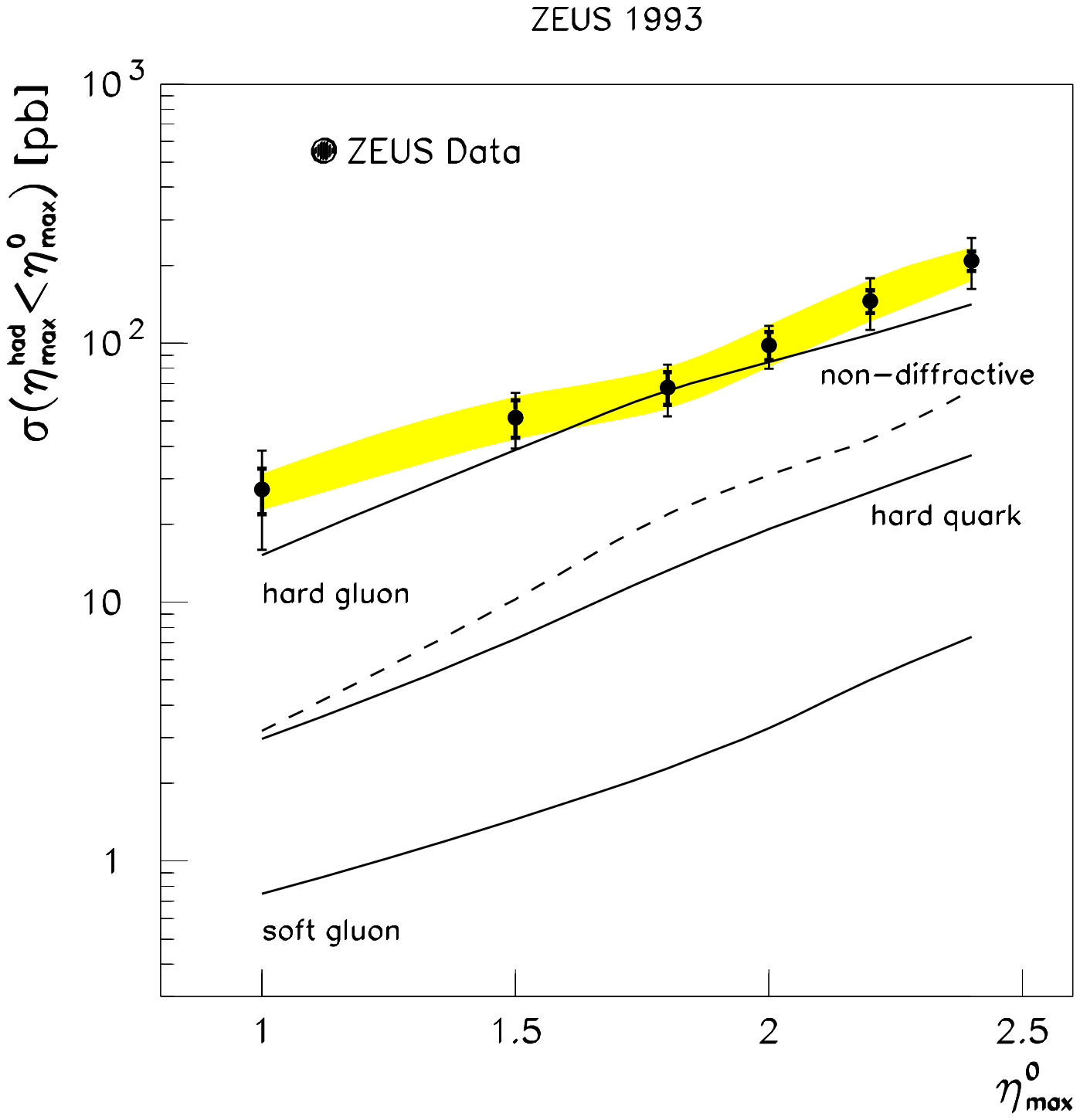}
\caption{\label{figdra2}{ Measured integrated $ep$ cross section
  $\sigma(\eta^{had}_{max}<\eta^0_{max})$ for inclusive jet production
  for $-1 < \eta^{jet}<1$ and $E_T^{jet} >$~8~GeV in the kinematic
  region $Q^2 \leq$~4~GeV$^2$ and $ 0.2 < y < 0.85$ (dots). The
  measurements are not corrected for the contributions from
  non-diffractive processes and double dissociation. The
  inner error bars represent the statistical errors of the data, and
  the total error bars show the statistical and systematic errors
  $-$not associated with the energy scale of the jets$-$ added in
  quadrature. The shaded band displays the uncertainty due to the
  energy scale of the jets. For comparison, PYTHIA and POMPYT
  calculations (for the same conditions as in Fig.~2)
  are included.}}
\end{figure}

 The results for $d\sigma/d\eta^{jet}(\eta^{had}_{max}<1.8)$ and
$\sigma(\eta^{had}_{max}<\eta^0_{max})$ are presented in
Figs.~\ref{figdra1} and~\ref{figdra2} and in Tables~\ref{table1}
and~\ref{table2}. The differential cross section is flat as a function
of $\eta^{jet}$. Since the measured jet cross sections refer
to events with a large rapidity gap they include a contribution from
double dissociation. The statistical errors of the
measurements are indicated as the inner error bars in
Figs.~\ref{figdra1} and \ref{figdra2}. They are $\sim 30$\% for
$d\sigma/d\eta^{jet}(\eta^{had}_{max}<1.8)$ and constitute the
dominant source of uncertainty. For
$\sigma(\eta^{had}_{max}<\eta^0_{max})$ the statistical error increases
from 8\% to 20\% as $\eta^0_{max}$ decreases. A detailed study of
the systematic uncertainties of the measurements has been carried out
\cite{zesep94,thesisk}. The sources of uncertainty include the
dependence on the choice of the parton densities in the pomeron, the
simulation of the trigger, the cuts used to select the data, and the
absolute energy scale of the $cal$ jets \cite{zesep94}.

 The following systematic uncertainties related to the
$\eta_{max}$-cut were studied: the $\eta^{cal}_{max}$ variable in the
data and the simulated events was recomputed after removing the
CAL cells with $\eta > 3.25$ in order to check the dependence on the
detailed simulation of the forward region of the detector, resulting in
changes up to 13\% for $\sigma(\eta^{had}_{max}<\eta^0_{max})$ and up
to 18\% for $d\sigma/d\eta^{jet}(\eta^{had}_{max}<1.8)$ (except at the
most forward data point, where the statistics are small and the change
amounts to 37\%); the energy threshold in the computation of
$\eta^{cal}_{max}$ for data and simulated events was decreased to
300~MeV, yielding changes up to 11\% for
$d\sigma/d\eta^{jet}(\eta^{had}_{max}<1.8)$ and up to
14\% for $\sigma(\eta^{had}_{max}<\eta^0_{max})$ (except at the most
backward data point, where the statistics are small and the change
amounts to 27\%).

 The dominant source of systematic error is the absolute energy scale
of the $cal$ jets, known to within 5\%, which results in a 20\% error.
The systematic uncertainties not associated with the energy scale of
the jets were added in quadrature to the statistical errors and are
shown as the total error bars. The additional uncertainty due to the
energy scale of the jets is shown as a shaded band. The systematic
uncertainties have large bin to bin correlations. They are to be
understood as a conservative estimate of the error associated with each
data point. An additional overall normalisation uncertainty
of 3.3\% from the luminosity determination is not included.

\subsection{Comparison to non-diffractive model predictions}

 The contribution to the measured cross sections from non-diffractive
processes was estimated using PYTHIA including resolved and direct
processes. $Had$ jets were selected in the generated\footnote{ These
generated events were analysed at the hadron level.} events using the
same jet algorithm as for the data and calculating $\eta^{had}_{max}$
as explained in section 6. In PYTHIA the occurrence of a rapidity gap
is exponentially suppressed and arises from a fluctuation in the
pseudorapidity distribution of the final state hadrons.
The calculations using MRSD$_-$ \cite{mrsd} for
the proton and GRV-HO \cite{grv} for the photon parton densities are
compared to the measurements in Figs.~\ref{figdra1} and \ref{figdra2}.
The non-diffractive contribution does not reproduce the
measurements. For the measured
$d\sigma/d\eta^{jet}(\eta^{had}_{max}<1.8)$ the non-diffractive
contribution is close to the data only at the most forward measured
point. For the remaining $\eta^{jet}$ range, the data are a factor
between $2$ and $7$ above the expectations from non-diffractive
processes. In the measured $\sigma(\eta^{had}_{max}<\eta^0_{max})$, the
non-diffractive contribution as predicted by PYTHIA is smaller than the
data by factors between $3$ and $9$. These comparisons, together with
the features of the data shown in the previous section, demonstrate
that the measured jet cross sections with a large rapidity gap cannot
be accounted for by non-diffractive processes. However, in the
discussion below, this non-diffractive contribution will be subtracted
from the data.

\subsection{Comparison to diffractive model predictions}

 The measured cross sections are compared to the predictions
of the models for diffractive hard scattering mediated by pomeron
exchange, as implemented in the POMPYT generator. The predictions have
been obtained by selecting $had$ jets in the generated events using the
same jet algorithm as for the data and calculating $\eta^{had}_{max}$
as explained in section 6.

 In a first step, the predictions of POMPYT, using the DL flux factor
and the parametrisations of the pomeron parton densities suggested
theoretically (see section 3.1), and assuming $\Sigma_{\Pma} = 1$, are
compared to the measured cross sections in Figs.~\ref{figdra1}
and~\ref{figdra2}. For this initial comparison, the contributions from
non-diffractive and double dissociation processes have not been taken
into account. As mentioned earlier (see section 3.1), the $\mu^2$
dependence of the parton densities has been neglected and, hence, the
argument of $\Sigma_{\Pma}(\mu^2)$ is omitted. The scale relevant for
the measured jet cross sections is $\mu^2 \sim (E_T^{jet})^2$.
We start by discussing the results for
$d\sigma/d\eta^{jet}(\eta^{had}_{max}<1.8)$. The shape of the
predictions of POMPYT using a hard parton density compares well with
the measured shape of the cross section. The shape predicted by POMPYT
using a soft gluon density does not describe the data. The calculations
based on a soft gluon density are smaller than the measurements by
factors between $20$ and $50$. This type of parametrisation was already
disfavoured by previous studies \cite{h1lrgp94}. The predictions using
a hard quark density are too small by factors between $3$ and $10$, but
those using a hard gluon density reproduce the measurements well.
The predictions based on the IS flux factor lead to similar
conclusions. For the integrated cross section
$\sigma(\eta^{had}_{max}<\eta^0_{max})$, the measured shape is in each
case described by the expectations from POMPYT, although the
normalisation is incorrect by a factor depending on the model.
A soft gluon (hard quark) density yields a prediction which is too
small by factors between $30$ and $60$ ($5$ and $10$). A hard gluon
density for the pomeron gives a good description of the data.
Based on the samples of events of POMPYT, the measurements are
sensitive to $\beta$ values above approximately 0.3. Therefore, the
data is not sensitive to a possible additional contribution due to
a soft parton component in the pomeron. The data do not rule out a
possible contribution from a super-hard parton component in the pomeron
$[15-17]$.

 In principle, other processes could contribute to jet production with
a large rapidity gap. For example, the proton may emit a $\pi^+$
(instead of a pomeron), $p_i \rightarrow n_f \pi^+$, and a partonic
constituent of the $\pi^+$ undergoes a hard interaction with the photon
or its constituents. The contribution from this reaction to the data is
expected to be small due to the power law decrease, $\sim W^{-4}$, for
pion exchange. Monte Carlo calculations using POMPYT confirm these
expectations.

 In a second step, the data were compared to the predictions of POMPYT
based on a pomeron consisting of both quarks and gluons but without
assuming $\Sigma_{\Pma}=1$. In addition, the contribution from
non-diffractive processes and from double dissociation to
the measured cross sections were taken into account. The
non-diffractive contribution as predicted by PYTHIA\footnote{These
calculations give a good description of the inclusive jet differential
cross sections
(without the large-rapidity-gap requirement) in the range
$-1<\eta^{jet}<1$ \cite{zesep94}.} was subtracted bin by bin from
the data. The contribution from double dissociation for
large-rapidity-gap events was estimated to be $(15 \pm 10)$\%
\cite{zelrgd95}. This contribution was assumed to be independent of
$\eta^{jet}$ and was also subtracted from the data. After the above
subtractions, the data were compared with the predictions of POMPYT
using the DL flux factor and allowing for a mixture of the hard
gluon ($6 \beta (1-\beta)$) and the hard quark ($\frac{6}{4} \beta
(1-\beta)$) densities in the pomeron: a fraction $c_g$ for hard gluons
and $c_q =1-c_g$ for hard quarks. The overall normalisation of the
POMPYT prediction was left as a free parameter: $\Sigma_{\Pma}$.
For this study, the contribution to the cross sections and to
$\Sigma_{\Pma}$ from possible soft gluon and soft quark components
has been neglected.

\begin{figure}
\epsfysize=18cm
\epsffile{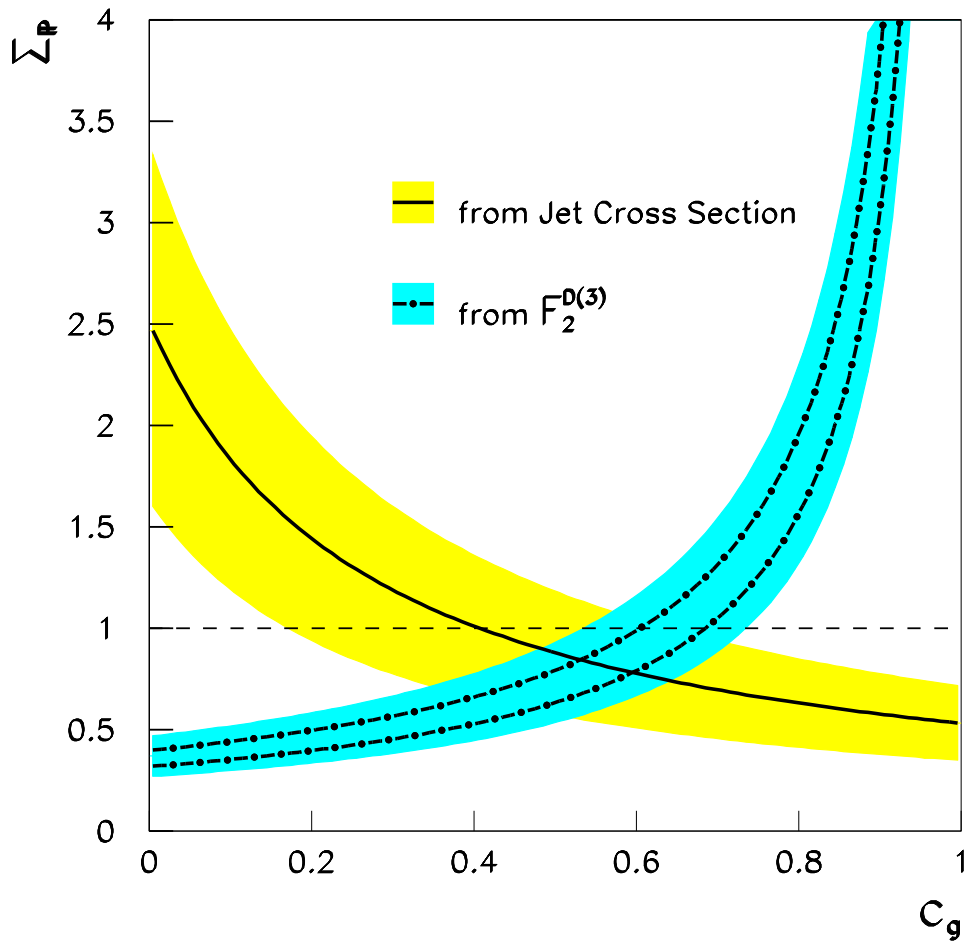}
\caption{\label{figdra4}{  The plane of the variables $\Sigma_{\Pma}$
  (momentum sum) and $c_g$ (relative contribution of hard
  gluons in the pomeron). The thick solid line displays the minimum for
  each value of $c_g$ obtained from the $\chi^2$ fit (the shaded area
  represents the $1$~$\sigma$ band around these minima) to the measured
  $d\sigma/d\eta^{jet}(\eta^{had}_{max}<1.8)$ using the
  predictions of POMPYT. The constraint imposed in the
  $\Sigma_{\Pma}-c_g$ plane by the measurement of the diffractive
  structure function in DIS ($F_2^{D(3)}$) $[10]$ for
  two choices of the number of flavours (upper dot-dashed line for
  $\Sigma_{{\Pma}q}=0.40$ and lower dot-dashed line for
  $\Sigma_{{\Pma}q}=0.32$) is also shown. The horizontal dashed line
  displays the relation $\Sigma_{\Pma}=1$.}}
\end{figure}

 For each value of $c_g$, a one-parameter ($\Sigma_{\Pma}$)
$\chi^2$-fit to the measured
$d\sigma/d\eta^{jet}(\eta^{max}_{had}<1.8)$ was performed. The results
are presented in Fig.~\ref{figdra4}. The thick solid line represents
the value of $\Sigma_{\Pma}$ for the minimum of the $\chi^2$-fit for
each value of $c_g$ and the shaded band represents the $1$~$\sigma$
range around those minima. For $c_g=1$ (gluons only) the fit yields
$\Sigma_{\Pma} = 0.5 \pm 0.2$ with $\chi^2_{min}=2.3$ for three degrees
of freedom, while for $c_g=0$ (quarks only) the fit yields
$\Sigma_{\Pma} = 2.5 \pm 0.9$ with $\chi^2_{min}=2.8$. The momentum sum
rule ($\Sigma_{\Pma} =1$) is approximately satisfied for
$0.2< c_g < 0.6$ (statistical errors only). Note that for this estimate
the DL form for the pomeron flux factor was assumed.

 This comparison of cross sections for jet production with a large
rapidity gap between data and model predictions is subject to the
following uncertainties:
\begin{itemize}
 \item The jet cross sections obtained from the Monte Carlo
       calculations presented here are leading order calculations. In
       these calculations, $\alpha_s(\mu^2)$  and the parton densities
       in the proton and the photon are evaluated at $\mu^2 =
       \hat{p}_T^2$. These computations may be affected by higher order
       QCD corrections, which are expected to change mainly the
       normalisation ($K$-factor). The agreement between the PYTHIA
       calculations of the inclusive jet differential cross sections
       and the measurements \cite{zesep94} indicate that in the case of
       the non-diffractive contribution the $K$-factor is close to $1$,
       within an uncertainty of $\sim 20$\%. The $K$-factor in the case
       of POMPYT is expected to be similar (with a similar
       uncertainty), as the same hard subprocesses are involved in the
       calculation of jet cross sections.
 \item The amount of the non-diffractive contribution to the
       measured cross section was modelled using PYTHIA with
       some choices for the proton and photon parton densities.
       This contribution is more sensitive to the choice of photon
       parton densities.
 \item The uncertainty in the estimation of the contribution
       from double dissociation.
 \item The POMPYT model for diffractive hard scattering assumes
       factorisation of the hard process with respect to the
       soft diffractive reaction. The extent to which this
       assumption is valid has to be determined experimentally
       through a detailed comparison of measurements for different
       reactions (see next section).
 \item The pomeron flux factors adopted in the various models are based
       on different assumptions for the $t$ and $x_{\Pma}$ dependences
       which are obtained from data on soft diffractive hadronic
       processes. The uncertainty in the procedure used to extract the
       flux is about 30\%.
\end{itemize}

 The differences between the results obtained in each of the studies
listed above and the central values were combined in quadrature to
yield the theoretical systematic uncertainties (not shown in
Fig.~\ref{figdra4}) of the fitted values of $\Sigma_{\Pma}$. These
uncertainties were then added in quadrature with the statistical and
systematic uncertainties of the measurements resulting in the following
ranges at the $1$~$\sigma$ level: $1.4<\Sigma_{\Pma}<3.8$ for
$c_g=0$ and $0.3<\Sigma_{\Pma}<0.9$ for $c_g=1$. The range in
$\Sigma_{\Pma}$ assumes the DL convention for the pomeron flux factor.
This normalisation has recently been discussed by Landshoff \cite{land}
who concludes that the normalisation is arbitrary up to a
multiplicative factor $A$. If the normalisation is changed by a factor
$A$, the range of the momentum sum is given by $1.4/A < \Sigma_{\Pma} <
3.8/A$ for $c_g=0$ and by $0.3/A<\Sigma_{\Pma}<0.9/A$ for $c_g=1$.

 In summary, the comparison of model predictions with the jet cross
section measurements favours those models where the partonic content of
the pomeron has a hard contribution. Given the uncertainties mentioned
above and the DL convention for the normalisation of the pomeron flux
factor, the data can be reproduced by a pomeron whose partonic content
varies between a pure hard quark density with momentum sum given by
$1.4<\Sigma_{\Pma}<3.8$ and a pure hard gluon density with
$0.3<\Sigma_{\Pma}<0.9$.

\subsection{The gluon content of the pomeron}

 The HERA experiments have recently presented the first measurements of
the diffractive structure function in DIS \cite{h1lrgd95,zelrgd95}.
The results show that the quark densities in the pomeron have a hard
and a soft contribution. Assuming the DL form for the pomeron flux
factor, the DIS data do not favour a pomeron structure function which
simultaneously fulfils the momentum sum rule and consists exclusively
of quarks.

 If the pomeron parton densities are universal and describe both
DIS and photoproduction processes, the DIS results together with the
photoproduction data further constrain the partonic content of
the pomeron. The measured diffractive structure function in DIS
($F_2^{D(3)}(\beta,Q^2,x_{\Pma})$) \cite{h1lrgd95,zelrgd95} can be
used to extract the contribution of the quarks to the momentum sum
($\Sigma_{{\Pma}q}(Q^2)$). The integral of $F_2^{D(3)}$ over $x_{\Pma}$
and $\beta$ is proportional to $\Sigma_{{\Pma}q}(Q^2)$:
\begin{equation}
  \int_{x_{{\Pma}{min}}}^{x_{{\Pma}{max}}} dx_{\Pma} \int_{0}^{1}
  d\beta \; F_2^{D(3)}(\beta,Q^2,x_{\Pma})
  = k_f \cdot  \Sigma_{{\Pma}q}(Q^2) \cdot I_{flux}
\end{equation}
where $I_{flux}$ is the integral of the pomeron flux factor over $t$
and over the same region in $x_{\Pma}$, and $k_f$ is a number which
depends on the number of flavours assumed ($5/18$ for two flavours and
$2/9$ for three flavours). For the left-hand side of Eq.~(9), the
parametrisation of $F_2^{D(3)}(\beta,Q^2,x_{\Pma})$ obtained in
\cite{zelrgd95} was used. The integral was performed over the range
$6.3 \cdot 10^{-4} < x_{\Pma} < 10^{-2}$ of the ZEUS DIS measurements.
The DL form for the pomeron flux factor was used to compute $I_{flux}$
for the right hand side of Eq.~(9). This procedure yields an estimate
of $\Sigma_{{\Pma}q}(Q^2)$: $0.32 \pm 0.05$ ($0.40\pm 0.07$) for two
(three) flavours. These estimates are based on a parametrisation of
$F_2^{D(3)}(\beta,Q^2,x_{\Pma})$ which was determined in the large
$\beta$ region ($0.1 < \beta < 0.8$) and is assumed to be valid for
the entire region $0 < \beta < 1$. In the range of
$Q^2$ where the DIS measurements were done, $8$~GeV$^2 < Q^2 <
100$~GeV$^2$, the pomeron structure function is approximately
independent of $Q^2$ and, thus, the estimated $\Sigma_{{\Pma}q}(Q^2)$
does not depend upon $Q^2$. It should be noted that the scales at which
the parton densities in the pomeron are probed in DIS, $Q^2$, and in
photoproduction, $\mu^2$, are comparable. The estimate from DIS imposes
a constraint on the $\Sigma_{\Pma}-c_g$ plane which, combined with the
estimates obtained in the preceding section, restricts the
allowed ranges for $\Sigma_{\Pma}$ and the relative contributions of
quarks and gluons ($c_g$). The DIS constraint, which can be written as
$\Sigma_{\Pma} \cdot (1-c_g)= 0.32$ ($0.40$) for the two choices of
the number of flavours, is included in Fig.~\ref{figdra4}
(the dark shaded area represents the uncertainty in this constraint).
Combining the estimates from photoproduction (thick solid line)
and DIS  yields $0.5 < \Sigma_{\Pma} < 1.1$ and $0.35< c_g < 0.7$
(statistical errors only).

 These results are subject to the uncertainties listed at the end of
section 7.3. The allowed range for $\Sigma_{\Pma}$ which
results from the combination of the DIS and photoproduction
measurements was evaluated for each source of systematic uncertainty.
Taking into account all the uncertainties mentioned, the comparison
between the DIS and photoproduction measurements gives $0.4 <
\Sigma_{\Pma} < 1.6$ for the momentum sum of the pomeron assuming the
DL convention for the flux. If the normalisation of the pomeron
flux factor is changed by a multiplicative factor $A$, the allowed
range of the momentum sum is given by $0.4/A <\Sigma_{\Pma} < 1.6/A$.

 It should be noted that the evaluation of the $c_g$ range allowed
by the DIS and photoproduction measurements is not affected by
the normalisation of the pomeron flux factor or the uncertainty on the
double dissociation contribution since they cancel out in the
comparison\footnote{This cancellation occurs as long as the same
pomeron flux factor is used in both DIS and photoproduction.}.
Taking into account the remaining uncertainties the combination of the
DIS and photoproduction data gives $0.3< c_g < 0.8$. This result does
not depend on the validity of the momentum sum rule for the pomeron.

\section{\bf Summary and conclusions}

 Measurements of $ep$ cross sections for inclusive jet photoproduction
with a large rapidity gap in $ep$ collisions at $\sqrt{s}=$ 296~GeV
using data collected by the ZEUS experiment in 1993 have been
presented. The measured jet cross sections are compared to perturbative
QCD calculations of diffractive hard processes and allow a model
dependent determination of the parton content of the pomeron. The
measurements require a contribution from a hard momentum density of the
partons in the pomeron. This result is consistent with the observations
of the UA8 Collaboration made in $p\bar{p}$ collisions. When the
measured jet cross sections are combined with the results on the
diffractive structure function in deep inelastic scattering at HERA,
first experimental evidence for the gluon content of the pomeron is
found. This evidence is independent of the normalisation of the flux of
pomerons from the proton and does not rely on assumptions on the
momentum sum of the pomeron. The data indicate that between 30\% and
80\% of the momentum of the pomeron carried by partons is due to hard
gluons.

\vspace{0.5cm}

\noindent {\Large\bf Acknowledgements}

\vspace{0.5cm}

 We thank the DESY Directorate for their strong support and
encouragement. The remarkable achievements of the HERA machine group
were essential for the successful completion of this work and are
greatly appreciated. We would like to thank J. Collins, G. Ingelman
and G. Kramer for valuable discussions.


\newpage

\begin{table}
\centering
\begin{tabular}{|c||c|}                \hline
$\eta^{jet}$ & $d\sigma/d\eta^{jet}(\eta^{had}_{max}<1.8)\pm
$stat.$\pm$syst.$\pm$syst.
                    $E^{jet}_T$-scale \\
range        &
(in pb)  \\ \hline\hline
           &                                   \\
(-1,-0.5)  & $ 44\pm 12\pm  8$ $^{+10}_{-8}$   \\
           &                                   \\
(-0.5,0)   & $ 39\pm 11\pm  5$ $^{+8}_{-5}$    \\
           &                                   \\
(0,0.5)    & $ 37\pm 10\pm 10$ $^{+8}_{-6}$    \\
           &                                   \\
(0.5,1)    & $ 16\pm  6\pm  7$ $^{+3}_{-3}$    \\
           &                                   \\ \hline
\end{tabular}
\caption{\label{table1}
{Measured differential $ep$ cross section
$d\sigma/d\eta^{jet}(\eta^{had}_{max}<1.8)$ for inclusive jet
production for $E_T^{jet} > $~8~GeV and in the kinematic region
$Q^2 \leq$~4~GeV$^2$ and $ 0.2 < y < 0.85$. The $\eta^{jet}$ ranges
used for the measurements are shown. The cross sections are given at
the centre of each $\eta^{jet}$ bin. The statistical and
systematic errors are also indicated. The systematic uncertainties
associated with the energy scale of the jets are quoted separately. The
overall normalisation uncertainty of 3.3\% is not included.}}
\end{table}

\begin{table}
\centering
\begin{tabular}{|c||c|}                \hline
$\eta_{max}^0$ &
$\sigma(\eta_{max}<\eta_{max}^0)\pm$stat.$\pm$syst.$\pm$syst.
                                     $E^{jet}_T$-scale \\
     &
(in pb)                                                \\ \hline\hline
     &                                                 \\
1.0  & $   27.3\pm   5.5\pm  9.9$ $^{+5.9}_{-4.5}$     \\
     &                                                 \\
1.5  & $   51.7\pm   8.5\pm  9.2$ $^{+10.7}_{-9.1}$    \\
     &                                                 \\
1.8  & $   67  \pm    10\pm   12$ $^{+13}_{-11}$       \\
     &                                                 \\
2.0  & $   98  \pm    12\pm   14$ $^{+20}_{-17}$       \\
     &                                                 \\
2.2  & $  145  \pm    15\pm   29$ $^{+30}_{-23}$       \\
     &                                                 \\
2.4  & $  208  \pm    18\pm   43$ $^{+40}_{-34}$       \\
     &                                                 \\ \hline
\end{tabular}
\caption{\label{table2}
{Measured integrated $ep$ cross section
$\sigma(\eta^{had}_{max}<\eta_{max}^0)$ for inclusive jet production
for $E_T^{jet} > $~8~GeV and $-1<\eta^{jet}<1$ in the kinematic region
$Q^2\leq$~4~GeV$^2$ and $ 0.2 < y < 0.85$. The statistical and
systematic errors are also indicated. The systematic uncertainties
associated with the energy scale of the jets are quoted separately. The
overall normalisation uncertainty of 3.3\% is not included.}}
\end{table}

\newpage
\clearpage

\parskip 0mm

\newpage
\clearpage

\parskip 0mm

\newpage
\clearpage

\parskip 0mm

\newpage
\clearpage

\parskip 0mm

\end{document}